\documentclass{PoS}
\usepackage{mathptmx}
\usepackage{calrsfs}
\usepackage{pdfpages}
\usepackage{cite}
\DeclareMathAlphabet{\mathcal}{OMS}{zplm}{m}{n}

\let\OLDthebibliography\thebibliography
\renewcommand\thebibliography[1]{
  \OLDthebibliography{#1}
  \setlength{\parskip}{0pt}
  \setlength{\itemsep}{-3pt}
\footnotesize
}

\title{Interglueball potential in $SU(N_c)$ lattice gauge theory}

\ShortTitle{Interglueball potential in $SU(N_c)$ lattice gauge theory}

\author{\speaker{Nodoka~Yamanaka}\\
        Yukawa Institute for Theoretical Physics, Kyoto University, Kitashirakawa-Oiwake, Kyoto 606-8502, Japan\\
        E-mail: \email{ynodoka@yukawa.kyoto-u.ac.jp}
}
\author{Hideaki Iida\\
        Department of Physics, The University of Tokyo, 7-3-1 Hongo, Bunkyo-ku, Tokyo 113-0033, Japan
}
\author{Atsushi Nakamura\\
        School of Biomedicine, Far Eastern Federal University, 690950 Vladivostok, Russia\\
        Research Center for Nuclear Physics, Osaka University, Ibaraki, Osaka 567-0047, Japan\\
        Theoretical Research Division, Nishina Center, RIKEN, Wako 351-0198, Japan
}
\author{Masayuki Wakayama\\
        Center for Extreme Nuclear Matters (CENuM), Korea University, Seoul 02841, Republic of Korea\\
        Department of Physics, Pukyong National University (PKNU), Busan 48513, Republic of Korea
}

\abstract{
We report on our calculation of the interglueball potentials in $SU(2)$, $SU(3)$, and $SU(4)$ lattice Yang-Mills theories using the indirect (so-called HAL QCD) method.
We use the cluster decomposition error reduction technique to improve the statistical accuracy of the glueball correlators.
After calculating the glueball scattering cross section in $SU(2)$ Yang-Mills theory and combining with the observational data of the dark matter mass distributions, we derive the lower limit on the scale parameter.
}

\FullConference{37th International Symposium on Lattice Field Theory - Lattice2019\\
		16-22 June 2019\\
		Wuhan, China}

\begin{document}

\section{Introduction}

\vspace{-0.5em}

The dark matter (DM) is representing a significant fraction of the energy content of the Universe, but we currently do not know the theory which is governing this sector \cite{Bertone:2004pz,Battaglieri:2017aum}.
Among many candidates of new physics beyond the standard model, we focus on the ``dark'' $SU(N_c)$ Yang-Mills theory (YMT) \cite{Soni:2016gzf,Kribs:2016cew} which has the advantage to have good naturalness thanks to the dimensional transmutation, thus avoiding important fine-tunings of massive parameters.

The lightest particle in the YMT is the 0$^{++}$ glueball, which respects the properties of the observed DM.
The glueballs are nonperturbative objects, so the extraction of their dynamical information absolutely requires the lattice calculation \cite{deForcrand:1984eeq,Teper:1987wt,Albanese:1987ds,Teper:1998kw,Morningstar:1999rf,Bali:2000vr,Lucini:2001ej,Ishii:2001zq,Ishii:2002ww,Lucini:2004my,Chen:2005mg,Lucini:2010nv,Gregory:2012hu,Lucini:2012gg,Yamanaka:2019aeq,Yamanaka:2019yek}.
An important quantity in the physics of DM is the self-scattering cross section \cite{Spergel:1999mh}, which is constrained by observational data such as the galactic collision.
The calculation of the interhadron scattering on lattice recently knew significant progress thanks to technical improvements.
Through this calculation, we expect to constrain the scale parameter of $SU(N_c)$ YMT.
In this proceedings contribution, we report on our calculation of the interglueball potentials in $SU(2)$, $SU(3)$, and $SU(4)$ YMTs using the indirect method (the so-called HAL QCD method) \cite{Ishii:2006ec,Aoki:2009ji,HALQCD:2012aa} and the cluster decomposition error reduction technique (CDERT) \cite{Liu:2017man}.
We also show the preliminary constraint on the scale parameter of $SU(2)$ YMT.

\vspace{-0.5em}

\section{Setup of the calculation}

\vspace{-0.5em}

In this work, we simulate $SU(2)$ ($\beta =2.5$), $SU(3)$ ($\beta =5.7$), and $SU(4)$ ($\beta =10.789$) YMTs on $16^3\times 24$ lattice with the standard plaquette action.
The configurations are generated with the pseudo-heat-bath method.
To derive physical quantities from lattice calculations, we have to set the scale.
However, we do not know the scale of the YMT since the dark matter particles have not been identified so far.
We therefore leave it as a free parameter $\Lambda$.
We note that all calculated quantities are finally expressed in the unit of $\Lambda$.

The relation between $\Lambda$ and the string tension $\sigma$ was fitted from the analysis of the running coupling \cite{Allton:2008ty,Teper:2009uf}, as
\begin{eqnarray}
\frac{\Lambda}{\sqrt{\sigma}}
=
0.503(2)(40)+ \frac{0.33(3)(3)}{N_c^2}
.
\label{eq:scalestringtension}
\end{eqnarray}
The string tension was calculated for several $N_c$ and $\beta$.
By combining the result of these calculations with Eq. (\ref{eq:scalestringtension}), we obtain the lattice spacing in terms of the scale parameter (see Table~\ref{table:lattice_spacing}).

\begin{table}[h]
\begin{center}
\begin{tabular}{cccc}
\hline \hline
$N_c$ & $\beta$ & $a \sqrt{\sigma}$ & $a$ $(\Lambda^{-1})$\\
\hline
2 & 2.5 & 0.1834 (26) \cite{Teper:1998kw} & 0.107 (8) \\
3 & 5.7 & 0.3933 (16) \cite{Lucini:2004my} & 0.212 (16) \\
4 & 10.789 & 0.2706 (8) \cite{Lucini:2004my} & 0.142 (3) \\
\hline
\end{tabular}
\end{center}
\vspace{-1em}
\caption{
Relation between the lattice spacing and the scale parameter $\Lambda$ for several $SU(N_c)$ YMTs.
The numbers in parenthesis denote the combined statistical and systematic errors.
}
\label{table:lattice_spacing}
\vspace{-0.5em}
\end{table}

\vspace{-0.5em}

\section{Glueball correlators on lattice and the HAL QCD method}

\vspace{-0.5em}

We define the $0^{++}$ glueball operator on lattice as
\begin{equation}
\hspace{-0.15em} 
\phi (t, \vec{x}) 
=
{\rm Re} [
P_{12} (t, \vec{x}) 
+P_{12} (t, \vec{x}+a\vec{e}_3) 
+P_{23} (t, \vec{x}) 
+P_{23} (t, \vec{x}+a\vec{e}_1) 
+P_{31} (t, \vec{x}) 
+P_{31} (t, \vec{x}+a\vec{e}_2) 
]
,
\label{eq:glueballop}
\end{equation}
where $P_{ij}$ ($i,j = 1,2,3$) are the plaquette operator in $i-j$ direction, with the unit vectors $a\vec{e}_{1,2,3}$.
We note that the $0^{++}$ glueball operator has a vacuum expectation value, so we have to subtract it in order to calculate physical correlators.

\begin{figure}[hbt]
\begin{center}
\includegraphics[width=.45\columnwidth]{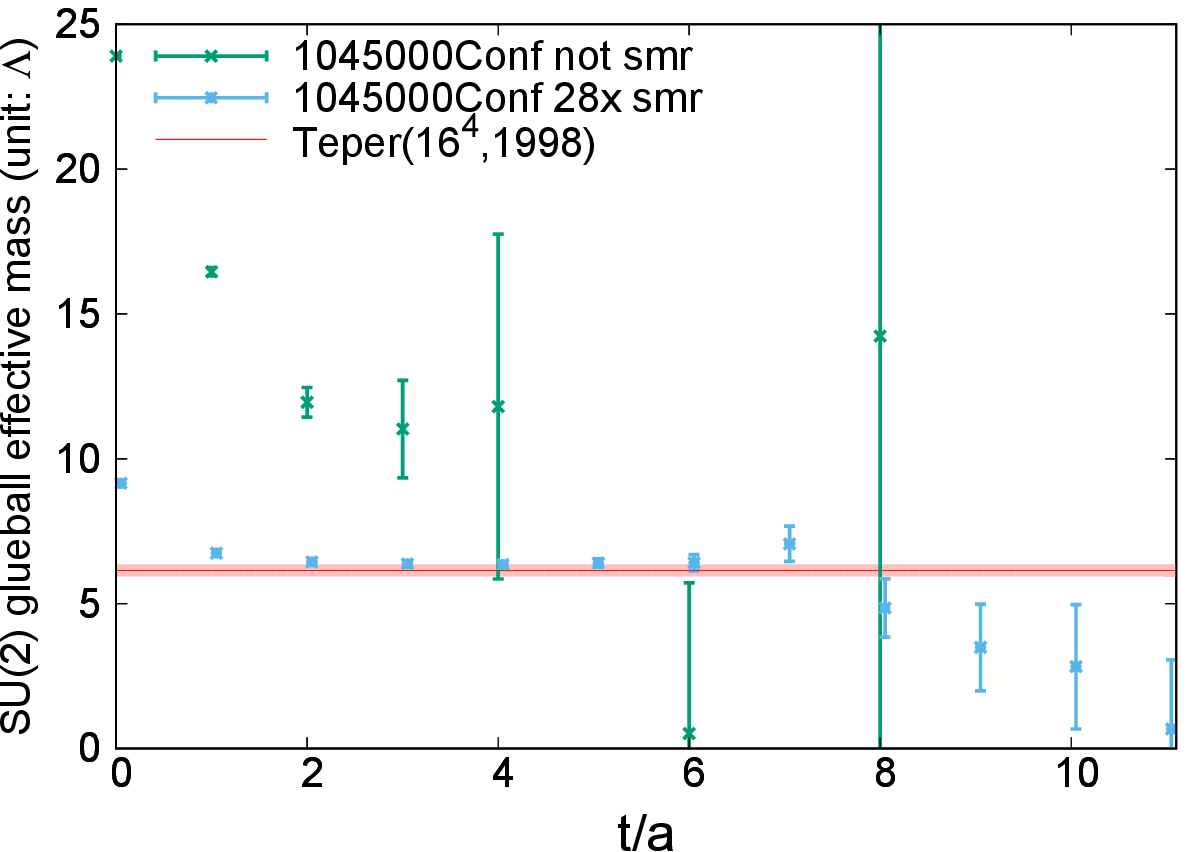}
\hspace{1em}
\includegraphics[width=.45\columnwidth]{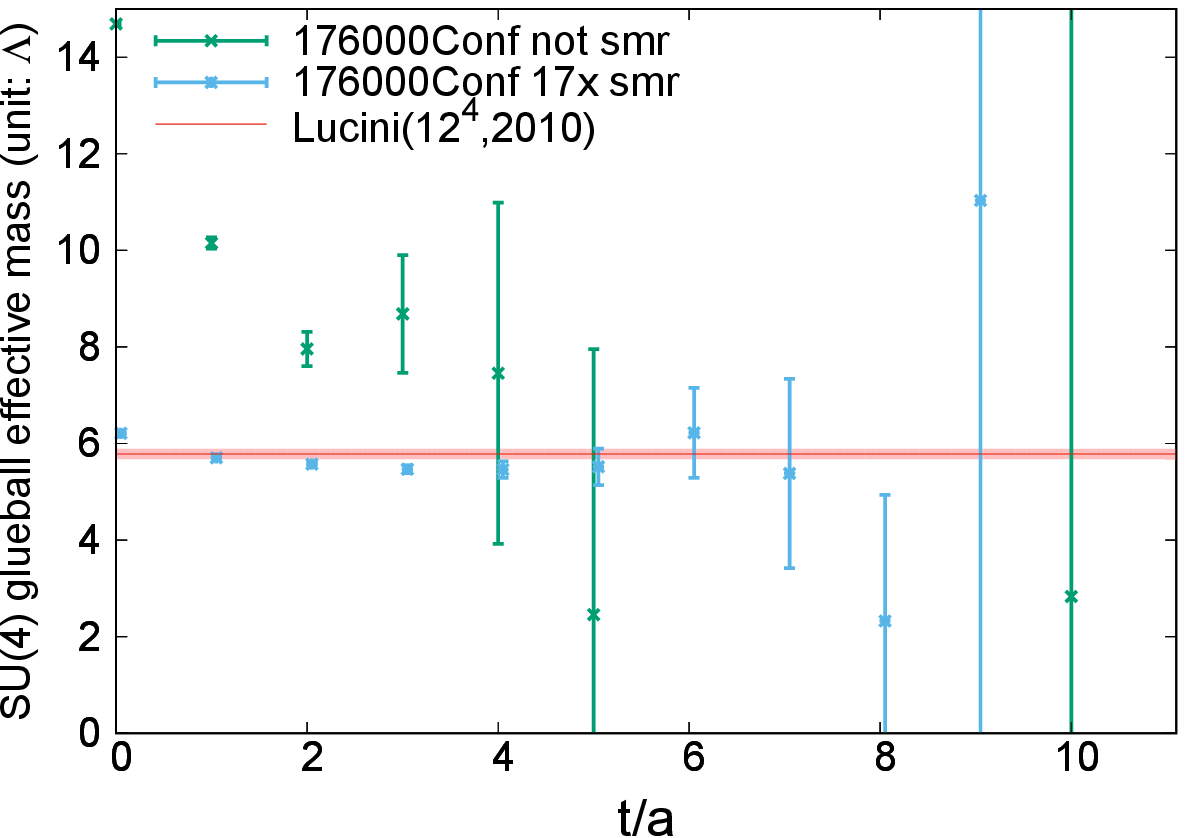}
\vspace{-0.5em}
\caption{
Effective mass plot of the 0$^{++}$ glueball two-point functions in $\beta = 2.5$ $SU(2)$ YMT (1045000 confs, left panel), and in $\beta = 10.789$ $SU(4)$ YMT (176000 confs, right panel) on $16^3 \times 24$ lattices.
}
\label{fig:su2_beta2p5_1045000_glueball_effmass}
\end{center}
\vspace{-1.5em}
\end{figure}

The glueball operator may be improved with the APE smearing \cite{Albanese:1987ds,Ishii:2001zq,Ishii:2002ww}.
It is constructed by maximizing 
\begin{equation}
{\rm Re\, Tr} [U_i^{(n+1)}(t, \vec{x}) V_i^{(n)\dagger}(t, \vec{x}) ]
, 
\label{eq:smearing}
\end{equation}
where $U_i^{(n)}$ is the link variable after $n$ iterations, and 
\begin{eqnarray}
V_i^{(n)}(t, \vec{x})
&\equiv &
\alpha U_i^{(n)}(t, \vec{x})
+\sum_{\pm j \neq i} U_j^{(n)}(t, \vec{x})
U_i^{(n)}(t, \vec{x}+\vec{e}_j) U_j^{(n)\dagger}(t, \vec{x}+\vec{e}_i)
.
\end{eqnarray}
We manually vary $\alpha $ and $n$ to find the optimized 0$^{++}$ glueball two-point function and its effective mass (see Fig. \ref{fig:su2_beta2p5_1045000_glueball_effmass}).
After optimization, we obtain the glueball masses (lattice unit) $m_\phi = 0.6857 (28)$ ($SU(2)$, $\beta =2.5$, 1045000 confs.), $m_\phi = 0.976 (10)$ ($SU(3)$, $\beta =5.7$, 158641 confs.), and $m_\phi = 0.776 (11)$ ($SU(4)$, $\beta =10.789$, 176000 confs.), consistent with previous works \cite{Teper:1998kw,Lucini:2010nv}.

\begin{figure}[htbp]
\centering
\includegraphics[width=0.45\textwidth,clip]{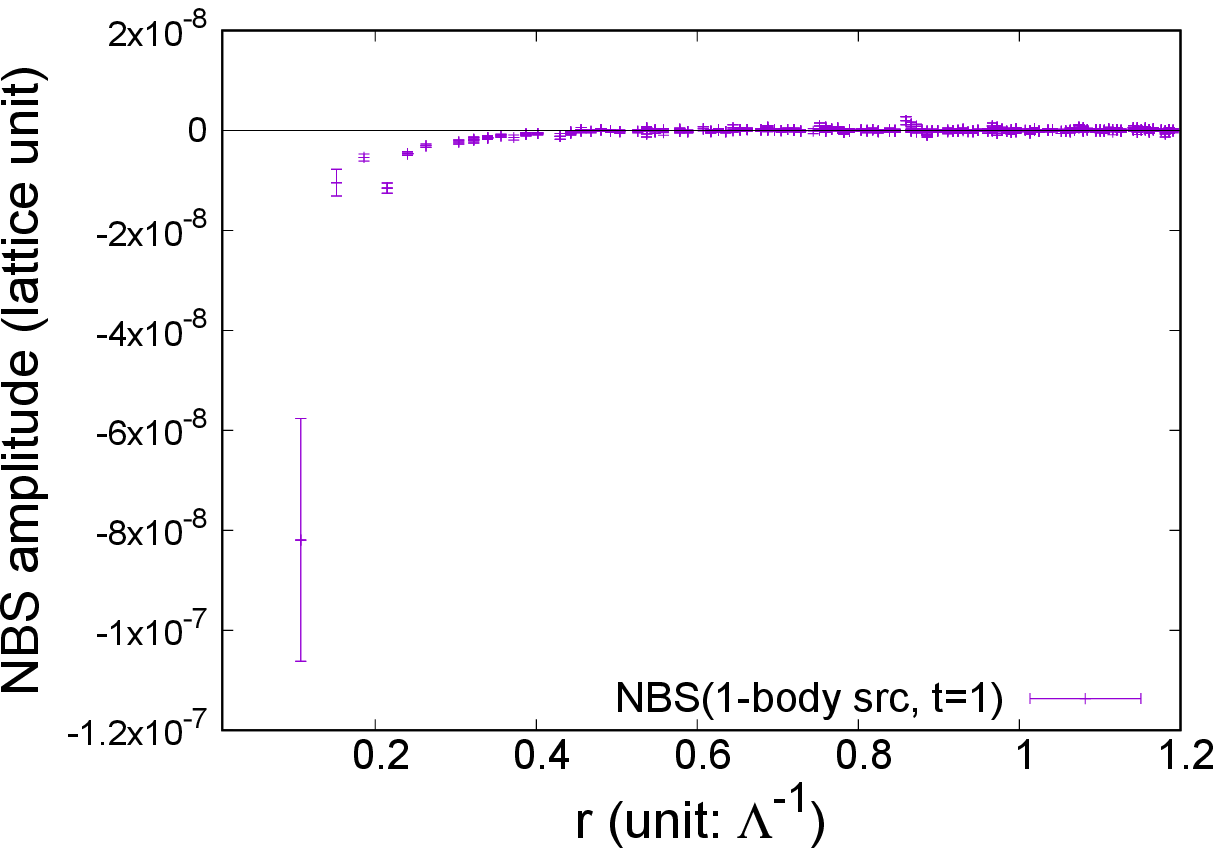}
\hspace{1em}
\includegraphics[width=0.45\textwidth,clip]{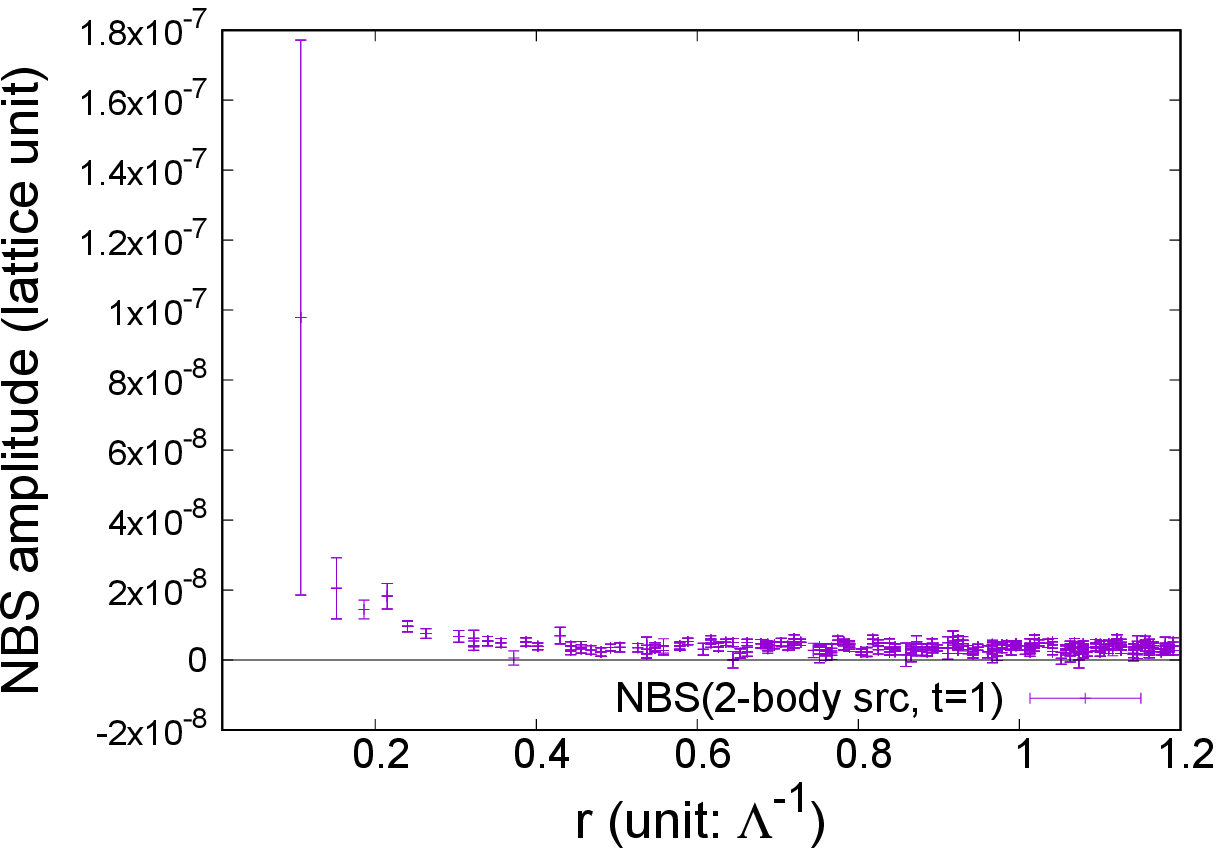}
\\
\vspace{0.5em}
\includegraphics[width=0.45\textwidth,clip]{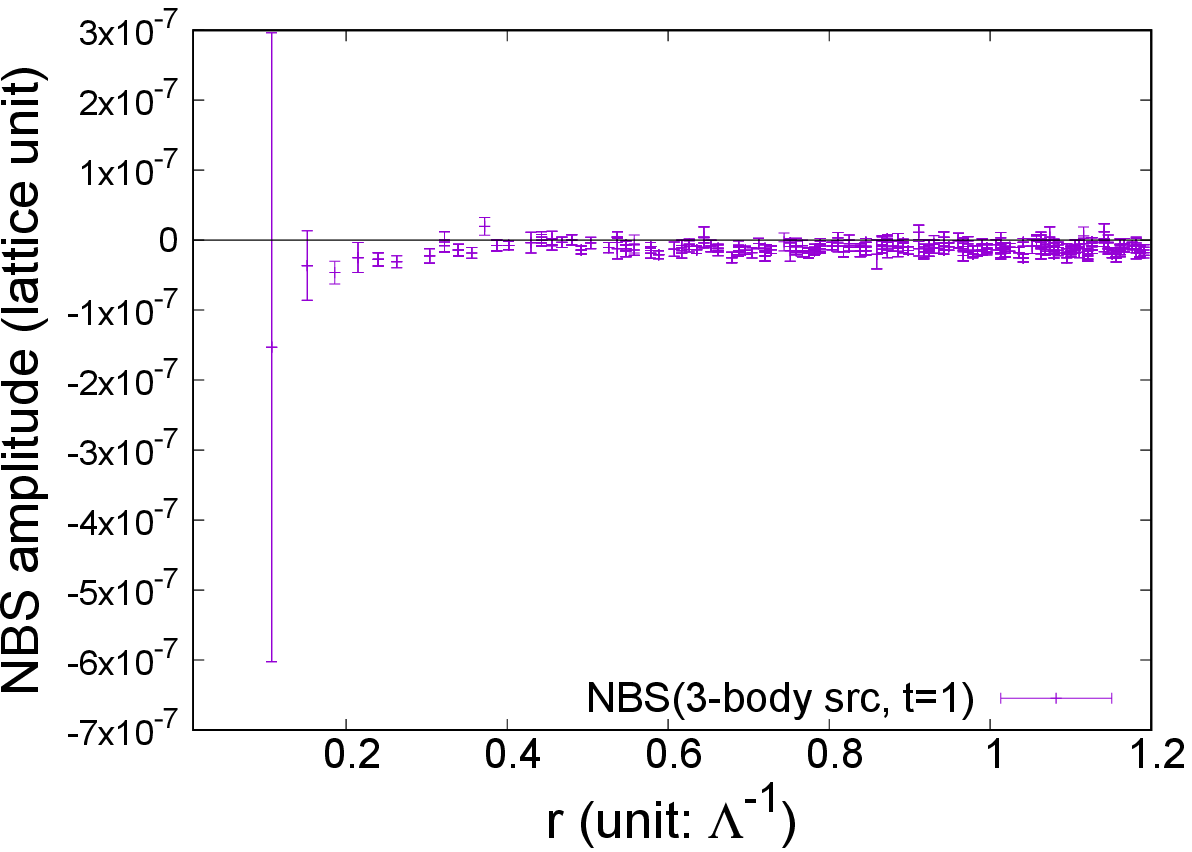}
\vspace{-0.5em}
\caption{\label{fig:su2_beta2p5_glueball_BS_n-body}
NBS amplitudes calculated with 1-, 2-, and 3-body wall sources in $SU(2)$ YMT (1045000 confs.).
The data were taken at the time slice $t=1$ ($t=0$ is the origin).
}
\vspace{-0.5em}
\end{figure}

The interglueball scattering is extracted from the following Nambu-Bethe-Salpeter (NBS) amplitude:
\begin{equation}
\Psi_{\phi \phi} 
(t,\vec{x}-\vec{y})
\equiv
\frac{1}{V} \sum_{\vec{r}} 
\langle 0 | T[\tilde \phi (t, \vec{x}+\vec{r})\tilde \phi (t, \vec{y}+\vec{r}) {\cal J}(0)] | 0 \rangle
.
\label{eq:NBS}
\end{equation}
Here ${\cal J}$ is the source operator with arbitrary power of 0$^{++}$ glueball operators.
This arbitrariness is due to the coupling of the two-glueball state with all other multi-glueball states.
We note that the multi-glueball operators have their own expectation values, so they must also be subtracted.
From the following relation
\begin{eqnarray}
&& \hspace{-2em}
\langle 0 | T[ [{\cal O}_{\rm snk} (t, \vec{r}) -\langle {\cal O}_{\rm snk} (t, \vec{r})\rangle ][{\cal O}_{\rm src} (0) -\langle {\cal O}_{\rm src} (0)\rangle ]  ] | 0 \rangle
\nonumber\\
&=&
\langle 0 | T[ [{\cal O}_{\rm snk} (t, \vec{r}) -\langle {\cal O}_{\rm snk} (t, \vec{r})\rangle ] {\cal O}_{\rm src} (0) | 0 \rangle
=
\langle 0 | T[ {\cal O}_{\rm snk} (t, \vec{r}) [{\cal O}_{\rm src} (0) -\langle {\cal O}_{\rm src} (0)\rangle ]  ] | 0 \rangle
, \ \ \ \ 
\end{eqnarray}
we see that it is sufficient to remove the expectation value of the source ${\cal J}$ to also subtract the one of the sink. 
We plot in Fig. \ref{fig:su2_beta2p5_glueball_BS_n-body} the NBS amplitude with the 1-, 2-, and 3-body sources in the $SU(2)$ YMT.
We see that they damp at long interglueball distance.
We also remark that for the cases with 2- and 3-body sources the NBS amplitudes are nonzero at long distance.
This is because the four-point (2-body source) and five-point (3-body source) functions can be cluster decomposed into nonzero correlator when the two sink operators are separated by a large distance, while the three-point (1-body source) correlator cannot.

\begin{figure}[htbp]
\vspace{-1.5em}
\centering
\includegraphics[width=0.45\textwidth,clip]{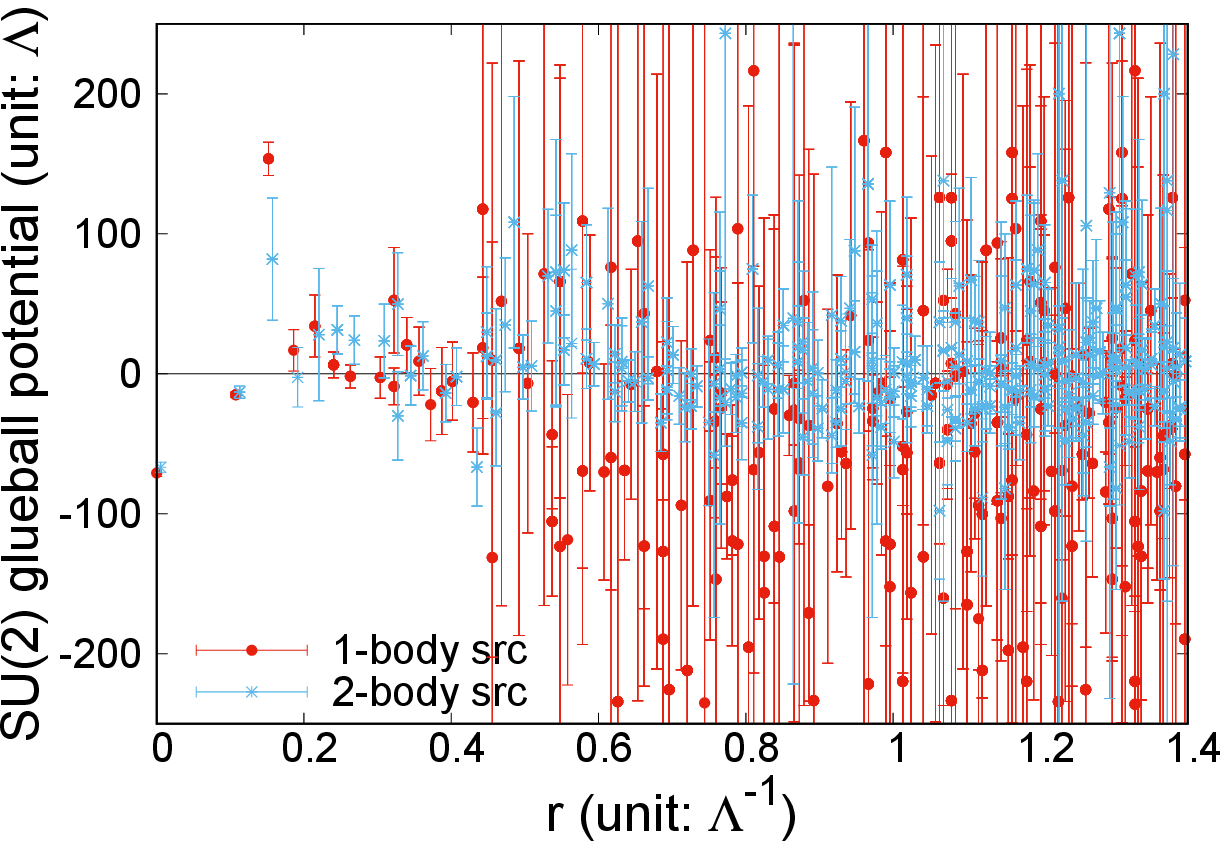}
\hspace{1em}
\includegraphics[width=0.45\textwidth,clip]{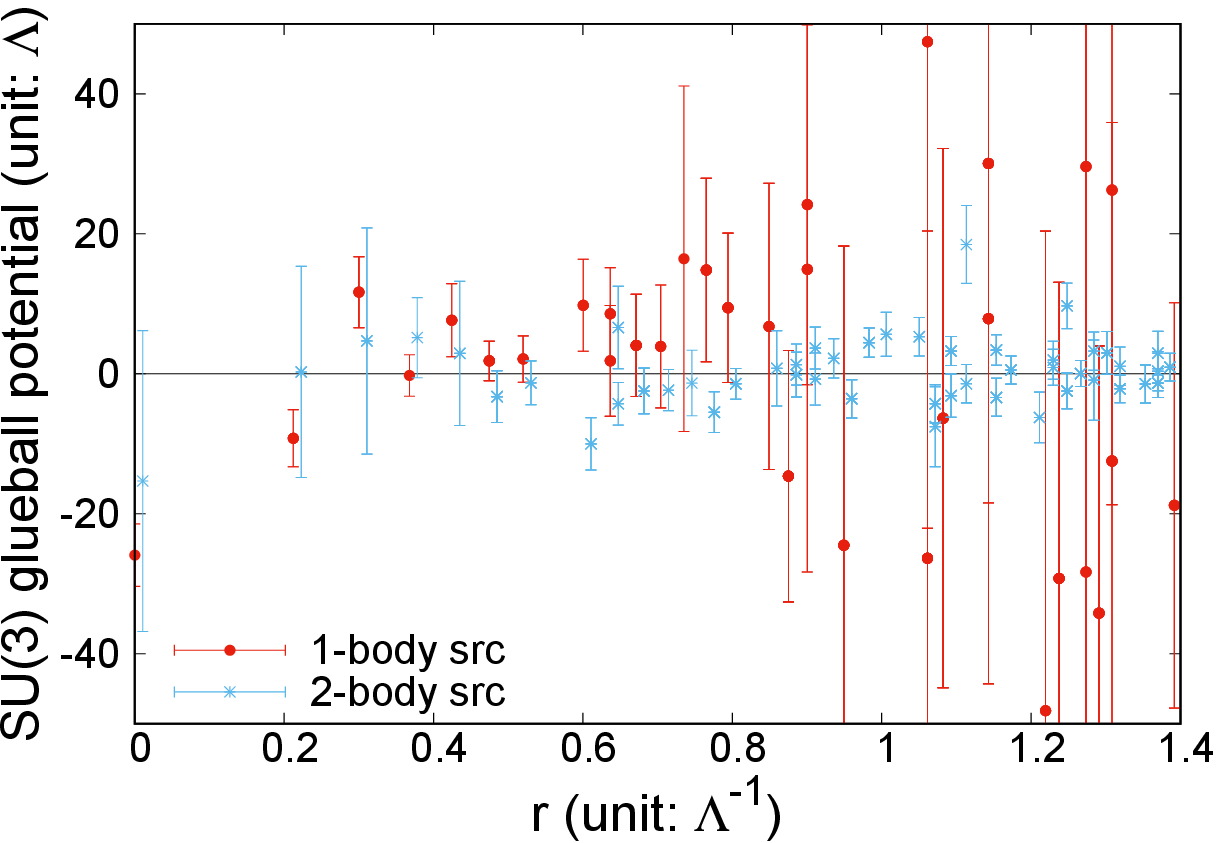}
\\
\vspace{0.5em}
\includegraphics[width=0.45\textwidth,clip]{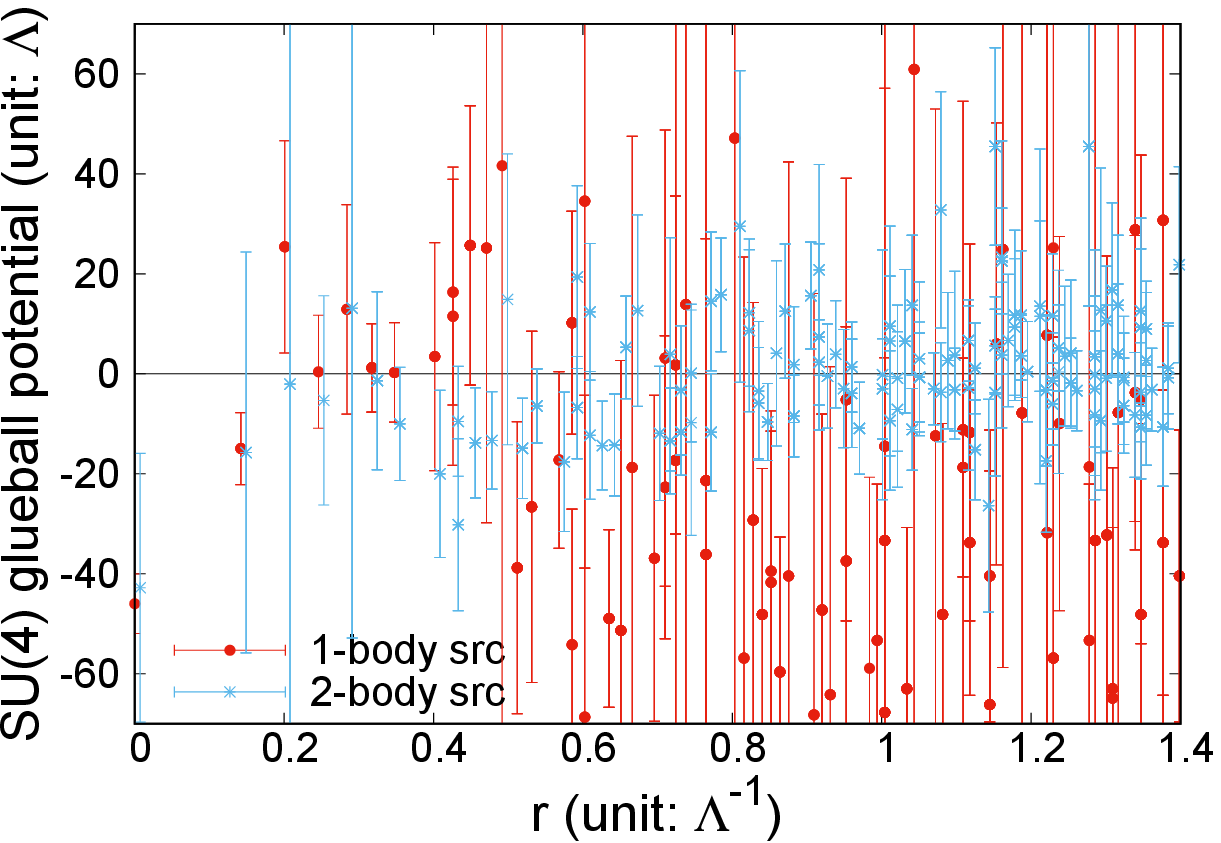}
\vspace{-0.5em}
\caption{\label{fig:glueball_potential_source_comparison}
Comparison of the 1-body and 2-body source calculations of the interglueball potential in $SU(2)$ ($\beta = 2.5$, 1045000 confs., upper left panel), $SU(3)$ ($\beta = 5.7$, 158641 confs., upper right panel), and  $SU(4)$ ($\beta = 10.789$, 176000 confs., lower panel) YMTs.
}
\end{figure}

In the indirect (HAL QCD) method, the physical scattering cross section is extracted from the interglueball potential.
It can be obtained by observing that the NBS amplitude fulfills the Schr\"{o}dinger equation \cite{Ishii:2006ec,Aoki:2009ji}.
Here we use the time-dependent formalism \cite{HALQCD:2012aa}
\begin{equation}
\Biggl[
\frac{1}{4m_\phi} \frac{\partial^2}{\partial t^2}-\frac{\partial}{\partial t} + \frac{1}{m_\phi} \nabla^2
\Biggr]
R(t,{\vec r})
=
\int d^3r' U({\vec r},{\vec r}')R(t,{\vec r}')
,
\label{eq:time-dependent_HAL}
\end{equation}
with $R(t,{\vec r}) \equiv \Psi_{\phi \phi} (t,{\vec r}) / e^{-2m_\phi t}$.
Here we take the local approximation $U({\vec r},{\vec r}') \approx V_{\phi \phi} (\vec{r}) \delta (\vec{r}-\vec{r}')$.
The important point of this approach is that we do not need to wait for the ground state saturation to obtain the potential.
Since Eq. (\ref{eq:time-dependent_HAL}) involves a second time derivative, we need three time slices to derive $V_{\phi \phi} (\vec{r})$.
We choose $t=1,2,3$ in our calculation ($t=0$ is the origin). 
In Fig. \ref{fig:glueball_potential_source_comparison}, we show the interglueball potential calculated in the indirect method.

\begin{figure}[hbt]
\begin{center}
\includegraphics[width=.45\columnwidth]{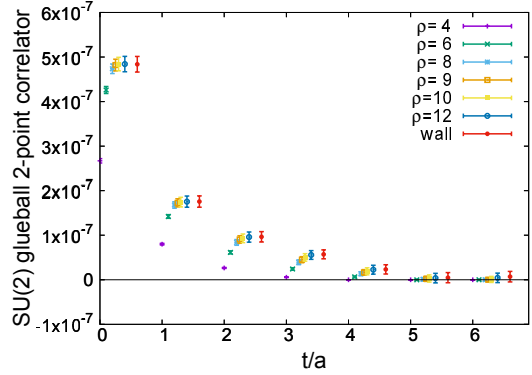}
\vspace{-0.5em}
\caption{
Glueball two-point function (smeared source and sink, 100 confs.) in $\beta =2.5$ $SU(2)$ YMT calculated with the CDERT.
We see that the correlator saturates at the cutoff $\rho=8$, and further increase of $\rho$ enlarges the statistical error.
}
\label{fig:glueball_2pt_cluster_decomp}
\end{center}
\vspace{-1.5em}
\end{figure}

The statistical error of glueball correlators is large due to the unsuppressed fluctuation of glueball operators at large space-time separation.
The CDERT \cite{Liu:2017man} can reduce such unwanted effect by applying a cutoff $\rho$ to the four-dimensional distance between glueball interpolating fields so as to not sum up the components of correlator which are cluster decomposed.
In Fig. \ref{fig:glueball_2pt_cluster_decomp}, we show the application of the CDERT to the glueball two-point function.
The correlator saturates at $\rho =8$ (lattice unit) for $SU(2)$ YMT with $\beta =2.5$.
This cutoff is optimal since larger cutoff enlarges the statistical error bar.

\begin{figure}[htbp]
\vspace{-2em}
\centering
\includegraphics[width=0.45\textwidth,clip]{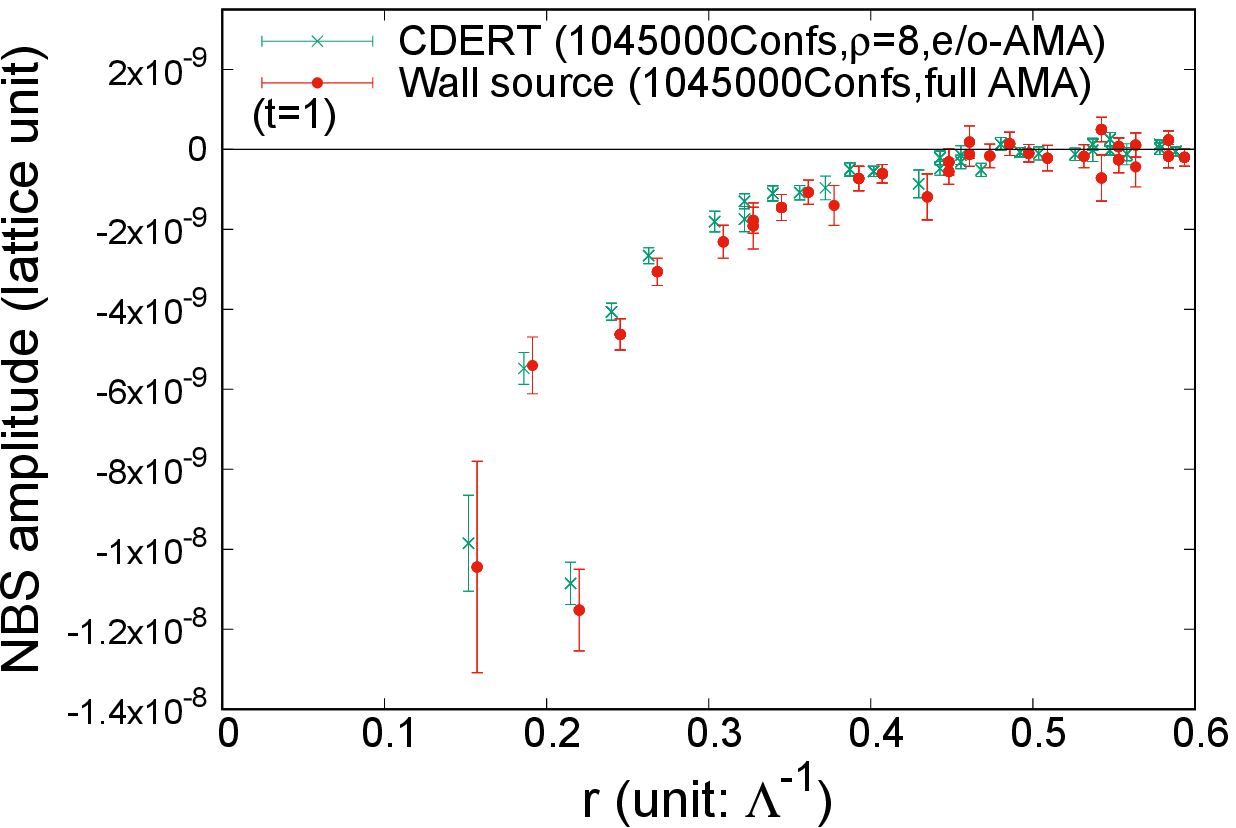}
\hspace{1em}
\includegraphics[width=0.45\textwidth,clip]{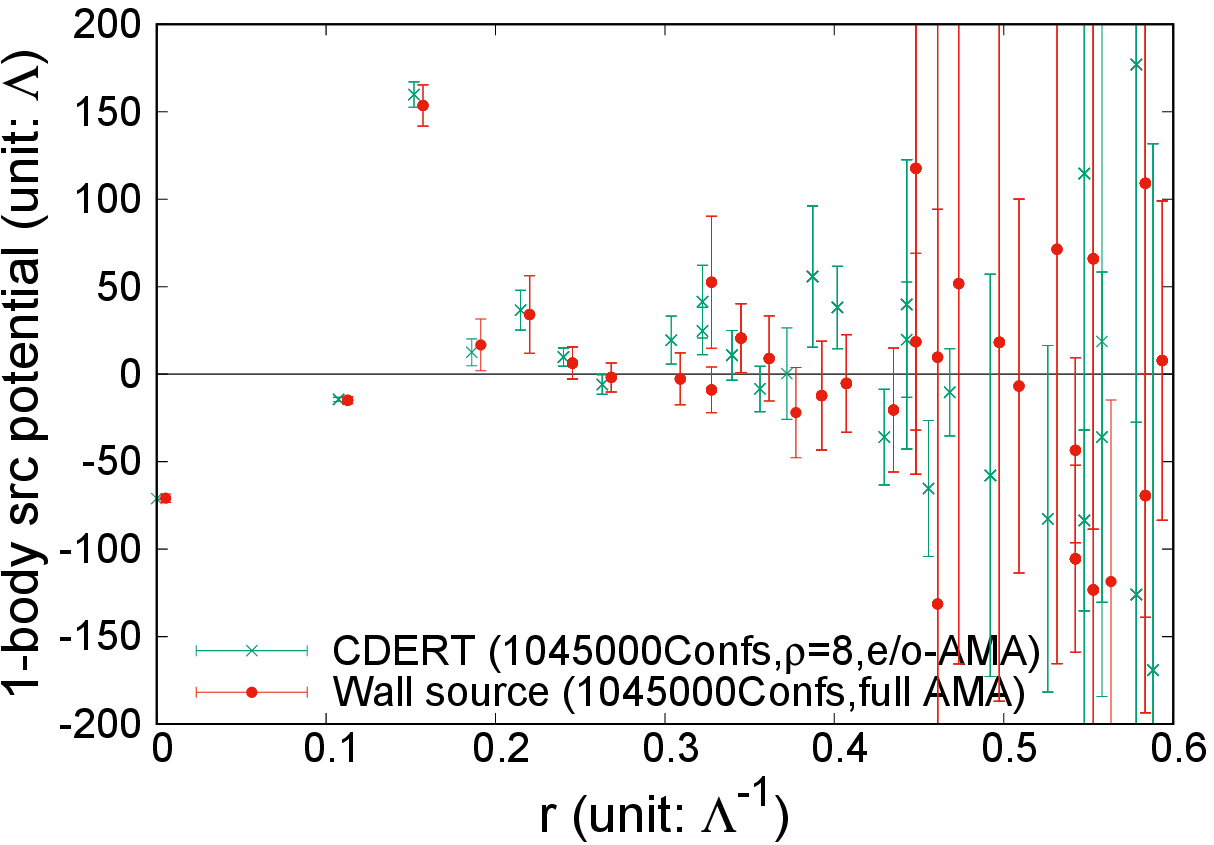}
\vspace{-0.5em}
\caption{\label{fig:glueball_BS_potential_cluster_decomp}
Comparison between the interglueball NBS amplitude and the potential ($\beta =2.5$ $SU(2)$ YMT) calculated in the CDERT with the cutoff $\rho =8$ and the wall source calculation.
The notation ``full AMA'' means that the NBS amplitude was averaged over all possible spatial translation, while ``e/o-AMA'' denotes that we took the average over even (odd) lattice for even (odd) time slices.
}
\vspace{-0.5em}
\end{figure}

We now apply the CDERT to the NBS amplitude with the 1-body source:
\begin{eqnarray}
\Psi'_{\phi \phi} (t,\vec{x}-\vec{y})
&=&
\frac{1}{V} \sum_{\vec{r}} 
\sum_{\vec r_{\rm src}\in C (t, \vec{x}+\vec{r} ) \bigcup C (t,\vec{y}+\vec{r} )} 
\langle 0 | T[\tilde \phi (t, \vec{x}+\vec{r})\tilde \phi (t, \vec{y}+\vec{r}) \tilde \phi (0, \vec r_{\rm src}  )] | 0 \rangle
.
\end{eqnarray}
Here $C (t, \vec{v} )$ is the projection of the four-dimensional hypersphere with the center $(t, \vec{v} )$ and with the radius (cutoff) $\rho$ onto the $t=0$ three-dimensional hyperplane.
The cutoffs are applied to the relative four-dimensional distances between the source operator and the sink ones.
In Fig. \ref{fig:glueball_BS_potential_cluster_decomp}, we compare the calculation of the NBS amplitude within the CDERT with that with the wall source.
We see that the CDERT is efficient in reducing the statistical error.

\section{Results}

\vspace{-0.5em}

\begin{figure}[hbt]
\begin{center}
\includegraphics[width=.45\columnwidth]{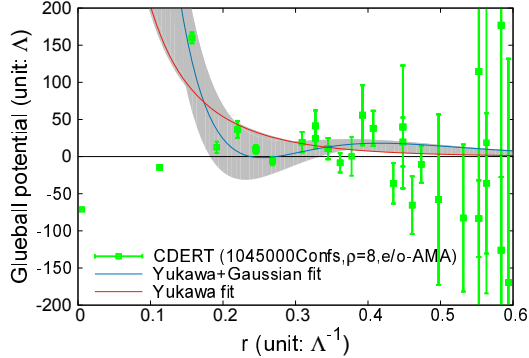}
\vspace{-0.5em}
\caption{
Fit of the 0$^{++}$ interglueball potential in $SU(2)$ YMT where the CDERT was applied with the cutoff $\rho =8$.
The notation ``e/o-AMA'' is the same as in Fig. \ref{fig:glueball_BS_potential_cluster_decomp}.
}
\label{fig:su2_beta2p5_1045000_glueball_potential_fit}
\end{center}
\vspace{-1.5em}
\end{figure}

After calculating the interglueball potential in $SU(2)$ YMT with the CDERT, we fit it with some adequate functionals to extract the scattering phase shift.
Here we use two fitting forms which depend on the glueball mass, namely $V_Y(r) = V_1 \frac{e^{-m_\phi r}}{r}$ and $V_{YG}(r) = V_1 \frac{e^{-m_\phi r}}{r}+V_2 e^{-\frac{(m_\phi r)^2}{2}}$.
We then obtain (in lattice unit) $V_Y(r) = 38.2 (2.1) \frac{e^{-m_\phi r}}{r}$ ($\chi^2/$d.o.f. = 12.6) and $V_{YG}(r) = 219.1 (15.1) \frac{e^{-m_\phi r}}{r} -68.2 (5.6) e^{-\frac{(m_\phi r)^2}{2}}$ ($\chi^2/$d.o.f. = 3.1), with the statistical error in parenthesis (see Fig. \ref{fig:su2_beta2p5_1045000_glueball_potential_fit}).

Now that we have the analytic form of the potential, we calculate the scattering phase shift and the cross section.
From the fitted potential, we calculate the scattering phase shift by solving the Schr\"{o}dinger equation:
\begin{eqnarray}
\biggl[
\frac{\partial^2}{\partial r^2} 
+k^2
-m_\phi V(r)
\biggr]
\psi (r)
=0
.
\end{eqnarray}
The solution of the above equation has the asymptotic behavior $\psi(r) \propto \sin [kr+\delta(k)]$, with the scattering phase shift $\delta(k)$.
Since the DM is nonrelativistic, $k$ may be considered as small, so the s-wave cross section is obtained as $\sigma_{\phi \phi} =\lim_{k\to 0}\frac{4 \pi}{k^2} \sin^2 [\delta(k)]$.
With the two fitting forms, we have $\sigma_{\phi \phi} = (3.2 - 3.4) \Lambda^{-2}$ $(V_Y \ {\rm fit})$ and $\sigma_{\phi \phi} = (6.7 - 7.1) \Lambda^{-2}$ $(V_{YG} \ {\rm fit})$.
The band denotes the statistical error.
Combining the above two, 
the interglueball scattering cross section in $SU(2)$ YMT is 
\begin{eqnarray}
\sigma_{\phi \phi} 
= 
(3.2 - 7.1) \Lambda^{-2}
\ \ \ {\rm (stat.+sys.)},
\label{eq:gblimit}
\end{eqnarray}
where the difference between the two fits were considered as the systematics.

Now that we have the relation between the cross section and $\Lambda$, the constraint on the scale parameter can be derived from observational data.
The comparison between the simulation and the observation of the shape of the galactic halo \cite{Rocha:2012jg} and galactic collisions \cite{Randall:2007ph} gives the limit $\sigma_{\rm DM} / m_{\rm DM} < 1 \, {\rm cm }^2 /$g.
By equating it with our result (\ref{eq:gblimit}), this yields a lower limit to the scale parameter of $SU(2)$ YMT $\Lambda > 50 \,{\rm MeV}$.
We do not calculate the constraints for $SU(3)$ and $SU(4)$ YMTs due to the large statistical error.
Instead, we estimate them according to the large $N_c$ argument, using the fact that the cross section scales as $1/N_c^4$.
We then have $\Lambda_{N_c} > \left( 2/N_c \right)^{\frac{4}{3}} \times 50 \,{\rm MeV}$.
We note that the correction of the $1/N_c$ expansion is $O(N_c^{-2})$, which is not small at $N_c=2$.

\vspace{-0.5em}

\section{Summary}

\vspace{-0.5em}

In this proceedings contribution, we reported on the result of our on-going calculation of the interglueball potential and cross section in the $SU(N_c)$ ($N_c= 2,3,4$) YMTs, which are good candidates to explain the DM physics.
Using the HAL QCD method, we could calculate the relation between the interglueball cross section and the scale parameter of $SU(2)$ YMT.
Combining with the observational data, we could obtain a lower limit on it.
To complete the study for all $N_c$ with sufficient accuracy, we definitely have to perform the same analysis for lower $N_c$, including the on-going calculations of $SU(3)$ and $SU(4)$ YMTs, which will provide us a better extrapolation.

This work was supported by ``Joint Usage/Research Center for Interdisciplinary Large-scale Information Infrastructures'' (JHPCN) in Japan (Project ID: jh180058-NAH).
The calculations were carried out on SX-ACE at RCNP/CMC of Osaka University.
MW was supported by the NRF grant funded by the Korea government (MSIT) (No.~2018R1A5A1025563).

\end{document}